\documentclass{article}%
\usepackage{amsmath}
\usepackage{amsfonts}
\usepackage{amssymb}
\usepackage{graphicx}%
\begin{document}

\title{ }

\begin{center}
{\LARGE Hamilton-Jacobi Approach for First Order Actions and Theories with
Higher Derivatives}
\end{center}

\bigskip

\begin{center}
M. C. Bertin$^{(a)}$\footnote{mcbertin@ift.unesp.br}, B. M. Pimentel$^{(a)}%
$\footnote{pimentel@ift.unesp.br}, P. J. Pompeia$^{(a,b)}$%
\footnote{pompeia@ift.unesp.br}

\bigskip

$^{(a)}$ Instituto de F\'{\i}sica Te\'{o}rica - Universidade Estadual
Paulista,\newline Rua Pamplona 145, 01405-900, S\~{a}o Paulo, SP, Brazil.

Telephone +55 11 3177-9029, FAX +55 11 3177-9080\newline$^{(b)}$ Comando-Geral
de Tecnologia Aeroespacial - Instituto de Fomento e Coordena\c{c}\~{a}o
Industrial - Divis\~{a}o de Confiabilidade Metrol\'{o}gica
Aeroespacial\newline Pra\c{c}a Mal. Eduardo Gomes, 50, 12228-901, S\~{a}o
Jos\'{e} dos Campos, SP, Brazil.\newline\bigskip

\textit{Abstract}
\end{center}

\textit{In this work we analyze systems described by Lagrangians with higher
order derivatives in the context of the Hamilton-Jacobi formalism for first
order actions. Two different approaches are studied here: the first one is
analogous to the description of theories with higher derivatives in the
hamiltonian formalism according to \cite{GitTyuLya, Gitman}; the second treats
the case where degenerate coordinate are present, in an analogy to reference
\cite{Degen}. Several examples are analyzed where a comparison between both
approaches is made.}

\bigskip

\begin{center}

\bigskip


\end{center}

\section{Introduction}

The interest of physicists by systems described by Lagrangians with
derivatives higher than one is not recent. Since 1850, when Ostrogradski
developed the first work concerning the hamiltonian formalism for systems with
higher derivatives \cite{Ostrog}, systems of this type have been used in many
relevant problems of Physics. As examples one can cite the works of Podolsky
\cite{Podolsky} and Bopp \cite{Bopp}, who independently proposed
generalizations of Electrodynamics containing second order derivatives, and
the works of Green \cite{Green}, who proposed a generalized meson-field
theory. Many other applications can be found in literature \cite{Stelle, West,
AAB, TG2ord}, where systems with higher derivatives have been successfully
used. When dealing with systems of this type, special attempts must be made
when the Lagrangian is singular.

As it is well known, the usual approach to deal with singular systems was
developed by Dirac in the early 1950`s \cite{Dirac0, Dirac2}, and its
application to system with higher derivatives was made in the 1980`s
\cite{GitTyuLya, Gitman}, which we refer henceforth as the standard approach.
Very recently, a new development for systems with higher derivatives and
degenerate coordinates, \textit{i.e.} coordinates whose derivatives are not
present in the theory (refered from now on as the degenerate coordinates
approach), was made by Gitman and Tyutin \cite{Degen} and a new definition of
singularity of a theory was proposed. According to the authors, this new
definition would be strictly correlated to the gauge character of the theory.
One interesting feature of both these developments lies on the fact that a
Lagrangian linear in the velocities can be written down.

Although widely accepted, Dirac formalism did not avoid the appearance of
other approaches, which always provide new points of view for the same
problems. One of them is the Hamilton-Jacobi (HJ) formalism, based on the
Carath\'{e}odory Equivalent Lagrangian method \cite{Carat}, whose approach to
singular systems was developed by G\"{u}ler \cite{Guler 1}, and since several
applications and extra developments have been made \cite{Rand Pim Jef, REF HJ,
REF HJ 2, REF HJ 3, REF HJ 4, coreanos}, including the study of systems with
second and higher order derivatives \cite{Rand Pim1, Rand Pim2}. An important
application in the context of this work was recently made in reference
\cite{HJ1ord}, where systems described by first order actions, \textit{i.e.}
Lagrangians linear in the velocities, were studied \textit{via} HJ formalism.
In \cite{HJ1ord} it was also shown how generalized brackets can be constructed
and how it is related to the existence of a symplectic structure in this formalism.

In this work we intend to analyze how systems with higher derivatives are
described in the HJ first order context. Considering that two different
approaches, the standard one and that with degenerate coordinates, were made
for systems with higher derivatives in the hamiltonian description, we intend
to compare how both of them can be described in the HJ formalism and see what
differences can be pointed out. For this in the next section we will make a
review of the HJ first order actions. In sequence we will apply this structure
for systems with higher order derivatives in analogy to the standard approach
(SA). Then the same will be made in the degenerate coordinates context (DC).
Several examples will be analyzed and the differences will be pointed out. At
last some remarks will be made.

\section{First Order Actions in HJ Formalism}

Let us consider the following Lagrangian%
\begin{equation}
L\left(  z_{B},\dot{z}_{B}\right)  =\dot{z}_{A}K^{A}\left(  z_{B}\right)
-V\left(  z_{B}\right)  ,~\ \ \ \ \ \ \ \ \ \ \ \ \ \ \ \ \ \ A,B=1,...,N.
\label{L_foa}%
\end{equation}
One then identifies the constraints%
\begin{align*}
\phi^{A}  &  \equiv p^{A}-K^{A}\left(  z\right)  =0,\\
\phi^{0}  &  \equiv p^{0}+V\left(  z\right)  =0.
\end{align*}
According to reference \cite{HJ1ord}, to verify the integrability conditions
of this system, one must analyze the matrix%
\begin{equation}
M^{AB}\equiv\left\{  \phi^{A},\phi^{B}\right\}  =\frac{\partial K^{B}%
}{\partial z_{A}}-\frac{\partial K^{A}}{\partial z_{B}}. \label{M_AB_foa}%
\end{equation}
If $\left\{  \phi^{A},\phi^{0}\right\}  =0,~M^{AB}=0$, then system is
integrable. When this is not the case, relations between the $z_{B}$ are
stablished, leading one to define generalized brackets. When the matrix
$M^{AB}$\ has rank $P=N-R$, then a submatrix $P\mathsf{\times}P$ exists such
that%
\[
\det\left(  M^{\dot{a}\dot{b}}\right)  \neq
0,\,\,\,\,\,\,\,\,\,\,\,\,\,\,\,\,\,\,\,\dot{a},\dot{b}=1,...,P,
\]
which means that $M_{\dot{a}\dot{b}}^{-1}$ also exists. The identification of
this submatrix separate $z_{B}$\ in two sets%
\[
z_{B}\rightarrow\left\{
\begin{array}
[c]{c}%
t_{\dot{b}}~,~\ \ \ \ \dot{b}=1,...,P,\\
t_{\beta}~,~\ \ \ \ \beta=1,...,R,
\end{array}
\right.
\]
where $t_{\dot{b}}$\ are the true dynamical variables and $t_{\beta}$\ play
the role of parameters in the theory. When this separation is done, some
integrability conditions must be satisfied:%
\begin{equation}
\left\{  \phi^{\alpha},\phi^{\beta}\right\}  =\left\{  \phi^{\alpha}%
,\phi^{\dot{b}}\right\}  M_{\dot{b}\dot{a}}^{-1}\left\{  \phi^{\dot{a}}%
,\phi^{\beta}\right\}  . \label{condint_foa}%
\end{equation}
Finally, generalized brackets can be constructed%
\begin{equation}
\left\{  F,G\right\}  _{\ast}\equiv\left\{  F,G\right\}  -\left\{
F,\phi^{\dot{b}}\right\}  M_{\dot{b}\dot{a}}^{-1}\left\{  \phi^{\dot{a}%
},G\right\}  , \label{PD_HJ_sing}%
\end{equation}
such that the differential of any function $E=E\left(  z\right)  $ becomes%
\begin{equation}
dE=\left\{  E,\phi^{\beta^{\prime}}\right\}  _{\ast}dt_{\beta^{\prime}}.
\label{dE_sing}%
\end{equation}
Hence the equations of motion can be obtained by setting $E=t_{\dot{b}}$.

\section{Theories with Higher Derivatives - Standard Approach}

In order to apply the structure of previous section to theories with higher
derivatives one can employ the development of references \cite{GitTyuLya,
Gitman}, which is described below. Let the system of interest be decribed by a
Lagrangian%
\begin{equation}
L=L\left(  q^{a},q^{\left(  l_{a}\right)  },t\right)  , \label{L}%
\end{equation}
where%
\[
q^{a\left(  l_{a}\right)  }\equiv\frac{d^{l_{a}}}{dt^{l_{a}}}q^{a}%
,\ \ \ \ \ \ \ \left(  a=1,...,n;l_{a}=1,...,N_{a}\right)  .
\]
This theory is constructed in a configuration space with $n$\ coordinates
$q^{a}$. Instead of studying the theory in this space, let one consider a
larger space with coordinates $x_{s}^{a},v^{a}$\ such that%
\begin{equation}
\left\{
\begin{array}
[c]{l}%
x_{s}^{a}\equiv q^{a\left(  s-1\right)  },\;\;s=1,...,N_{a},\left(
q^{a\left(  0\right)  }\equiv q^{a}\right) \\
v^{a}\equiv q^{a\left(  N_{a}\right)  }.
\end{array}
\right.  \label{newcoord}%
\end{equation}
With these definitions it can be stablished the following relations between
some coordinates and some of their time derivatives
\begin{equation}
\left\{
\begin{array}
[c]{c}%
x_{s+1}^{a}=\dot{x}_{s}^{a}\Rightarrow\dot{x}_{s}^{a}-x_{s+1}^{a}%
=0,\;\;\;s=1,...,N_{a}-1,\\
v^{a}=\dot{x}_{N_{a}}^{a}\Rightarrow\dot{x}_{N_{a}}^{a}-v^{a}=0.
\end{array}
\right.  \label{relations}%
\end{equation}
In this enlarged space the system can be described by the following Lagrangian%
\[
L^{v}\equiv L\left(  q^{a\left(  s-1\right)  }=x_{s}^{a},q^{a\left(
N_{a}\right)  }=v^{a},t\right)  =L^{v}\left(  x_{s}^{a},v^{a},t\right)  .
\]

Since the relations (\ref{relations}) must be satisfied, one can use Lagrange
multipliers to incorporate them in the theory:%
\begin{align}
\bar{L}^{v}  &  =\bar{L}^{v}\left(  x_{s}^{a},v^{a},t;\pi_{a}^{s}\right)
\equiv L^{v}\left(  x_{s}^{a},v^{a},t\right)  +\nonumber\\
&  +\sum_{a}\sum_{s=1}^{N_{a}-1}\pi_{a}^{s}\left(  \dot{x}_{s}^{a}-x_{s+1}%
^{a}\right)  +\sum_{a}\pi_{a}^{N_{a}}\left(  \dot{x}_{N_{a}}^{a}-v^{a}\right)
. \label{Lvbar}%
\end{align}
It must be noticed that, with these multipliers, the theory is now constructed
in larger space with coordinates $x_{s}^{a},v^{a},\pi_{a}^{s}$. $\bar{L}^{v}$
can be rewritten as a first order Lagrangian
\begin{align}
\bar{L}^{v}  &  =\sum_{a}\sum_{s=1}^{N_{a}-1}\pi_{a}^{s}\dot{x}_{s}^{a}%
+\sum_{a}\pi_{a}^{N_{a}}\dot{x}_{N_{a}}^{a}-H^{v}\left(  x_{s}^{a},v^{a}%
,t;\pi_{a}^{s}\right)  ,\label{Lvbar1}\\
H^{v}\left(  x_{s}^{a},v^{a},t;\pi_{a}^{s}\right)   &  \equiv\sum_{a}%
\sum_{s=1}^{N_{a}-1}\pi_{a}^{s}x_{s+1}^{a}+\sum\pi_{a}^{N_{a}}v^{a}%
-L^{v}\left(  x_{s}^{a},v^{a},t\right)  .\nonumber
\end{align}
Now the procedure of the previous section can be applied

\subsection{HJ first order approach}

In (\ref{Lvbar1}) the functions $K^{A}$\ and the constraints can be identified
as%
\begin{align*}
\left\{
\begin{array}
[c]{c}%
K_{a}^{v}=0\\
K_{a}^{x_{s}}=\pi_{a}^{s}\\
K_{\pi_{s}}^{a}=0
\end{array}
\right.   &  \Rightarrow\left\{
\begin{array}
[c]{c}%
\phi_{a}^{v}=p_{a}^{v}=0\\
\phi_{a}^{x_{s}}=p_{a}^{x_{s}}-\pi_{a}^{s}=0\\
\phi_{\pi_{s}}^{a}=p_{\pi_{s}}^{a}=0
\end{array}
\right.  ,\;s=1,...,N_{a},\\
\phi^{t}  &  =p^{0}+H^{v}=0,
\end{align*}
which allow one to construct the matrix $M^{AB}$:%
\[
M^{AB}=\frac{\partial K^{B}}{\partial z_{A}}-\frac{\partial K^{A}}{\partial
z_{B}},
\]%
\begin{align*}
\left(  M^{AB}\right)   &  =\left(
\begin{array}
[c]{ccc}%
0 & \frac{\partial K_{b}^{x_{s}}}{\partial v^{a}}-\frac{\partial K_{a}^{v}%
}{\partial x_{s}^{b}} & \frac{\partial K_{\pi_{s^{\prime}}}^{c}}{\partial
v^{a}}-\frac{\partial K_{a}^{v}}{\partial\pi_{c}^{s^{\prime}}}\\
\frac{\partial K_{a}^{v}}{\partial x_{s}^{b}}-\frac{\partial K_{b}^{x_{s}}%
}{\partial v^{a}} & 0 & \frac{\partial K_{\pi_{s^{\prime}}}^{c}}{\partial
x_{s}^{b}}-\frac{\partial K_{b}^{x_{s}}}{\partial\pi_{c}^{s^{\prime}}}\\
\frac{\partial K_{a}^{v}}{\partial\pi_{c}^{s}}-\frac{\partial K_{\pi_{s}}^{c}%
}{\partial v^{a}} & \frac{\partial K_{b}^{x_{s}}}{\partial\pi_{c}^{s^{\prime}%
}}-\frac{\partial K_{\pi_{s^{\prime}}}^{c}}{\partial x_{s}^{b}} & 0
\end{array}
\right)  =\\
&  =\left(
\begin{array}
[c]{ccc}%
0 & 0 & 0\\
0 & 0 & -\delta_{s^{\prime}}^{s}\delta_{b}^{c}\\
0 & \delta_{s^{\prime}}^{s}\delta_{b}^{c} & 0
\end{array}
\right)  .
\end{align*}

This matrix is singular, but there is an inversible submatrix%
\[
\left(  M^{\dot{c}\dot{b}}\right)  =\left(
\begin{array}
[c]{cc}%
0 & -\delta_{s^{\prime}}^{s}\delta_{b}^{c}\\
\delta_{s^{\prime}}^{s}\delta_{b}^{c} & 0
\end{array}
\right)  ,
\]%
\[
\left(  M_{\dot{a}\dot{c}}^{-1}\right)  =\left(
\begin{array}
[c]{cc}%
0 & \delta_{s^{\prime}}^{s}\delta_{c}^{a}\\
-\delta_{s^{\prime}}^{s}\delta_{c}^{a} & 0
\end{array}
\right)  ,
\]
showing that the variables can be separated in two sets:%
\begin{align*}
t_{\dot{a}}  &  =\left\{  x_{s}^{a};\pi_{a}^{s}\right\}  ,\\
t_{\beta}  &  =\left\{  v^{a}\right\}  .
\end{align*}

With this separation one must stablish the integrability conditions%
\[
\left\{  \phi^{\beta},\phi^{t}\right\}  =\left\{  \phi^{\beta},\phi^{\dot{a}%
}\right\}  M_{\dot{a}\dot{b}}^{-1}\left\{  \phi^{\dot{b}},\phi^{t}\right\}
,~\ \ \ \ \ \ \ \ \ \ \ \ \ \ \ \ \ \phi^{\dot{a}}=\left\{  \phi_{a}^{x_{s}%
};\phi_{\pi_{s}}^{a}\right\}  ,~\phi^{\beta}=\left\{  \phi_{a}^{v}\right\}  .
\]
Since $\left\{  \phi^{\dot{b}},\phi_{a}^{v}\right\}  =0$, as it can be
verified in the matrix $M^{AB}$, it follows%
\begin{align}
\left\{  \phi^{\beta},\phi^{t}\right\}   &  =0\Rightarrow\left\{  \phi_{b}%
^{v},\phi^{t}\right\}  =-\frac{\partial\phi^{t}}{\partial v^{b}}%
=\frac{\partial H^{v}}{\partial v^{b}}=0\Rightarrow\nonumber\\
&  \Rightarrow\pi_{a}^{N_{a}}-\frac{\partial L^{v}}{\partial v^{a}}=0.
\label{pi_na}%
\end{align}
Notice that these conditions stablish some of the Lagrange multipliers.

The next step is to stablish the Generalized Brackets, which allows one to
evaluate the total differential of any function $E=E\left(  x_{s}^{a}%
,v^{a},\pi_{a}^{s}\right)  $. For this it is necessary to know the following
quantities $\left\{  \phi^{\dot{a}},\phi^{\beta}\right\}  $\ and $\left\{
\phi^{\dot{a}},\phi^{t}\right\}  $. As mentioned before, the former are null
and only the last must be specified. First it mus be noticed that, for any
function $E=E\left(  x_{s}^{a},v^{a},\pi_{a}^{s}\right)  $, one has:%
\begin{align*}
\left\{  \phi_{a}^{x_{s}},E\right\}   &  =-\frac{\partial E}{\partial
x_{s}^{a}},\\
\left\{  \phi_{\pi_{s}}^{a},E\right\}   &  =-\frac{\partial E}{\partial\pi
_{a}^{s}}.
\end{align*}
For $E=\phi^{t}$\ it follows%
\[
\left\{  \phi^{\dot{a}},\phi^{t}\right\}  \rightarrow\left\{
\begin{array}
[c]{c}%
\left\{  \phi_{a}^{x_{s}},\phi^{t}\right\}  =-\frac{\partial\phi^{t}}{\partial
x_{s}^{aa}}\\
\left\{  \phi_{\pi_{s}}^{a},\phi^{t}\right\}  =-\frac{\partial\phi^{t}%
}{\partial\pi_{a}^{s}}%
\end{array}
\right.  .
\]

This way%
\[
dE=\left\{  E,\phi^{\beta^{\prime}}\right\}  _{\ast}dt_{\beta^{\prime}%
}=\left\{  E,\phi^{t}\right\}  _{\ast}dt+\left\{  E,\phi_{a}^{v}\right\}
_{\ast}dv^{a},
\]%
\begin{align*}
dE  &  =\left\{  E,\phi^{t}\right\}  dt-\left\{  E,\phi^{\dot{a}}\right\}
M_{\dot{a}\dot{b}}^{-1}\left\{  \phi^{\dot{b}},\phi^{t}\right\}  dt+\\
&  +\left\{  E,\phi_{a}^{v}\right\}  dv^{a}-\left\{  E,\phi^{\dot{a}}\right\}
M_{\dot{a}\dot{b}}^{-1}\left\{  \phi^{\dot{b}},\phi_{a}^{v}\right\}  dv^{a}%
\end{align*}%
\begin{equation}
dE=\frac{\partial E}{\partial t}dt+\frac{\partial E}{\partial v^{a}}%
dv^{a}+\left[  \frac{\partial E}{\partial x_{s}^{c}}\frac{\partial\phi^{t}%
}{\partial\pi_{c}^{s}}-\frac{\partial\phi^{t}}{\partial x_{s}^{c}}%
\frac{\partial E}{\partial\pi_{c}^{s}}\right]  dt. \label{dE}%
\end{equation}

Now the equations of motion can be obtained%
\begin{equation}
dt_{\dot{a}}=\frac{\partial t_{\dot{a}}}{\partial t}dt+\frac{\partial
t_{\dot{a}}}{\partial v^{a}}dv^{a}+\left[  \frac{\partial t_{\dot{a}}%
}{\partial x_{s}^{c}}\frac{\partial\phi^{t}}{\partial\pi_{c}^{s}}%
-\frac{\partial\phi^{t}}{\partial x_{s}^{c}}\frac{\partial t_{\dot{a}}%
}{\partial\pi_{c}^{s}}\right]  dt\Rightarrow dx_{s}^{a}=\frac{\partial H^{v}%
}{\partial\pi_{a}^{s}}dt \label{dx}%
\end{equation}%
\begin{equation}
d\pi_{a}^{s}=-\frac{\partial\phi^{t}}{\partial x_{s}^{a}}dt=-\frac{\partial
H^{v}}{\partial x_{s}^{a}}dt\Rightarrow\left\{
\begin{array}
[c]{l}%
d\pi_{a}^{s}=-\left(  \pi_{a}^{s-1}-\frac{\partial L^{v}}{\partial x_{s}^{a}%
}\right)  dt,\;\;s=2,...,N_{a},\\
d\pi_{a}^{1}=\frac{\partial L^{v}}{\partial x_{1}^{a}}dt.
\end{array}
\right.  \label{dpi}%
\end{equation}
Equations (\ref{dx}) are consistent with (\ref{relations}), while from
(\ref{dpi}) it can be seen that the other Lagrange multipliers are determined:%

\[
\pi_{a}^{s-1}=-\dot{\pi}_{a}^{s}+\frac{\partial L^{v}}{\partial x_{s}^{a}%
},\;\;s=2,...,N_{a}.
\]

\section{Degenerate Coordinates Approach}

Also with the intention to study the HJ approach for systems with degenerate
coordinates, one can start following the proposal made by Gitman and Tyutin in
reference \cite{Degen}, in which a first order Lagrangian is proposed.

Let one consider a Lagrangian%
\[
L=L\left(  q^{a},q^{\left(  l_{a}\right)  },t\right)  ,
\]
where%
\[
q^{\left(  l_{a}\right)  }\equiv\frac{d^{l_{a}}}{dt^{l_{a}}}q^{a}%
,\ \ \ \ \ \ \ \left(  a=1,...,n;l_{a}=1,...,N_{a}\right)  .
\]
The coordinates $q^{a}$ will be separated in two sets, $q^{\bar{a}}$ and
$q^{\tilde{a}}$, where $q^{\bar{a}}$ are coordinates whose derivatives are not
present in the Lagrangian, \textit{i.e.} degenerate coordinates, and
$q^{\tilde{a}}$\ are those whose derivatives of order $N_{\tilde{a}}$\ are
manifest in $L$, \textit{i.e.}%
\[
a=\left(  \bar{a},\tilde{a}\right)  ,\;\;\;\;N_{\bar{a}}=0;\;N_{\tilde{a}}%
\geq1.
\]

One then defines new coordinates%
\begin{align}
v^{\bar{a}}  &  \equiv q^{\bar{a}},\nonumber\\
x_{1}^{\tilde{a}}  &  \equiv q^{\tilde{a}},\nonumber\\
x_{s}^{\tilde{a}}  &  \equiv q^{\tilde{a}\left(  s-1\right)  }%
,\;\;s=2,...,N_{\tilde{a}},\nonumber\\
v^{\tilde{a}}  &  \equiv q^{\tilde{a}\left(  N_{\tilde{a}}\right)  },
\label{newcoord2}%
\end{align}
from where the following relations are identified:%
\begin{align}
x_{s+1}^{\tilde{a}}  &  =\dot{x}_{s}^{\tilde{a}}\Rightarrow\dot{x}_{s}%
^{\tilde{a}}-x_{s+1}^{\tilde{a}}=0,\;\;\;s=1,...,N_{\tilde{a}}-1,\nonumber\\
v^{\tilde{a}}  &  =\dot{x}_{N_{\tilde{a}}}^{\tilde{a}}\Rightarrow\dot
{x}_{N_{\tilde{a}}}^{\tilde{a}}-v^{\tilde{a}}=0. \label{relations2}%
\end{align}

With these new coordinates $L$ becomes%
\[
L^{v}\equiv L\left(  q^{\bar{a}}=v^{\bar{a}},q^{\tilde{a}\left(  s-1\right)
}=x_{s}^{\tilde{a}},q^{\tilde{a}\left(  N_{\tilde{a}}\right)  }=v^{\tilde{a}%
},t\right)  =L^{v}\left(  x_{s}^{\tilde{a}},v^{a},t\right)  .
\]
From this expression it is possible to define a new Lagrangian, $\bar{L}^{v}$,
in an extended space with coordinates $\left\{  x_{s}^{\tilde{a}},v^{a}%
,\pi_{\tilde{a}}^{s}\;\left(  s=1,...,N_{\tilde{a}}\right)  \right\}  $, which
differs from $L^{v}$\ by the presence of Lagrange multipliers, $\pi_{\tilde
{a}}^{s}$:%
\begin{align*}
\bar{L}^{v}  &  =\bar{L}^{v}\left(  x_{s}^{\tilde{a}},v^{a},t;\pi_{\tilde{a}%
}^{s}\right)  \equiv L^{v}\left(  x_{s}^{\tilde{a}},v^{a},t\right)  +\\
&  +\sum_{\tilde{a}}\sum_{s=1}^{N_{\tilde{a}}-1}\pi_{\tilde{a}}^{s}\left(
\dot{x}_{s}^{\tilde{a}}-x_{s+1}^{\tilde{a}}\right)  +\sum\pi_{\tilde{a}%
}^{N_{\tilde{a}}}\left(  \dot{x}_{N_{\tilde{a}}}^{\tilde{a}}-v^{\tilde{a}%
}\right)  .
\end{align*}
$\bar{L}^{v}$ is a first order Lagrangian, what becomes clearer when it is
rewritten as $L=\dot{z}_{A}K^{A}-V\left(  z_{A}\right)  $, where
$z_{A}=\left(  x_{s}^{\tilde{a}},v^{a};\pi_{\tilde{a}}^{s}\right)
,\;s=1,...,N_{\tilde{a}}$:%
\begin{align}
\bar{L}^{v}  &  =\sum_{\tilde{a}}\sum_{s=1}^{N_{\tilde{a}}-1}\pi_{\tilde{a}%
}^{s}\dot{x}_{s}^{\tilde{a}}+\sum\pi_{\tilde{a}}^{N_{\tilde{a}}}\dot
{x}_{N_{\tilde{a}}}^{\tilde{a}}-H^{v}\left(  x_{s}^{\tilde{a}},v^{a}%
,t;\pi_{\tilde{a}}^{s}\right)  ,\label{Lvbar2}\\
H^{v}\left(  x_{s}^{\tilde{a}},v^{a},t;\pi_{\tilde{a}}^{s}\right)   &
\equiv\sum_{\tilde{a}}\sum_{s=1}^{N_{\tilde{a}}-1}\pi_{\tilde{a}}^{s}%
x_{s+1}^{\tilde{a}}+\sum\pi_{\tilde{a}}^{N_{\tilde{a}}}v^{\tilde{a}}%
-L^{v}\left(  x_{s}^{\tilde{a}},v^{a},t\right)  . \label{Hv2}%
\end{align}
The procedure of reference \cite{HJ1ord} now can be applied.

It is important to notice that the main difference between the standard
approach and this one lies on the definition of the velocities of the
degenerate coordinates as the coordinates themselves in this last development.
When no degenerate coordinates exist, both approaches coincide; when they have
presence in $L$, the definitions of the velocities of degenerate coordinates
are different and one can verify that the space of the first approach is
larger than the space of the second.

\subsection{HJ first order approach}

The first step is to identify the functions $K^{A}$\ and the constraints. It
follows%
\begin{align*}
\left\{
\begin{array}
[c]{c}%
K_{a}^{v}=0\\
K_{\tilde{a}}^{x_{s}}=\pi_{\tilde{a}}^{s}\\
K_{\pi_{s}}^{\tilde{a}}=0
\end{array}
\right.   &  \Rightarrow\left\{
\begin{array}
[c]{c}%
\phi_{a}^{v}=p_{a}^{v}=0\\
\phi_{\tilde{a}}^{x_{s}}=p_{\tilde{a}}^{x_{s}}-\pi_{\tilde{a}}^{s}=0\\
\phi_{\pi_{s}}^{\tilde{a}}=p_{\pi_{s}}^{\tilde{a}}=0
\end{array}
\right.  ,\;s=1,...,N_{\tilde{a}},\\
\phi^{t}  &  =p^{0}+H^{v}=0.
\end{align*}
The next step is to compute the matrix $M^{AB}$:%
\[
M^{AB}=\frac{\partial K^{B}}{\partial z_{A}}-\frac{\partial K^{A}}{\partial
z_{B}},
\]%
\begin{align*}
\left(  M^{AB}\right)   &  =\left(
\begin{array}
[c]{ccc}%
0 & \frac{\partial K_{\tilde{b}}^{x_{s}}}{\partial v^{a}}-\frac{\partial
K_{a}^{v}}{\partial x_{s}^{\tilde{b}}} & \frac{\partial K_{\pi_{s^{\prime}}%
}^{\tilde{a}}}{\partial v^{a}}-\frac{\partial K_{a}^{v}}{\partial\pi
_{\tilde{a}}^{s^{\prime}}}\\
\frac{\partial K_{a}^{v}}{\partial x_{s}^{\tilde{b}}}-\frac{\partial
K_{\tilde{b}}^{x_{s}}}{\partial v^{a}} & 0 & \frac{\partial K_{\pi_{s^{\prime
}}}^{\tilde{a}}}{\partial x_{s}^{\tilde{b}}}-\frac{\partial K_{\tilde{b}%
}^{x_{s}}}{\partial\pi_{\tilde{a}}^{s^{\prime}}}\\
\frac{\partial K_{a}^{v}}{\partial\pi_{\tilde{a}}^{s}}-\frac{\partial
K_{\pi_{s}}^{\tilde{a}}}{\partial v^{a}} & \frac{\partial K_{\tilde{b}}%
^{x_{s}}}{\partial\pi_{\tilde{a}}^{s^{\prime}}}-\frac{\partial K_{\pi
_{s^{\prime}}}^{\tilde{a}}}{\partial x_{s}^{\tilde{b}}} & 0
\end{array}
\right)  =\\
&  =\left(
\begin{array}
[c]{ccc}%
0 & 0 & 0\\
0 & 0 & -\delta_{s^{\prime}}^{s}\delta_{\tilde{b}}^{\tilde{a}}\\
0 & \delta_{s^{\prime}}^{s}\delta_{\tilde{b}}^{\tilde{a}} & 0
\end{array}
\right)  .
\end{align*}
This matrix is singular with an inversible submatrix $M^{\dot{a}\dot{b}}$:%
\[
\left(  M^{\dot{a}\dot{b}}\right)  =\left(
\begin{array}
[c]{cc}%
0 & -\delta_{s^{\prime}}^{s}\delta_{\tilde{b}}^{\tilde{a}}\\
\delta_{s^{\prime}}^{s}\delta_{\tilde{b}}^{\tilde{a}} & 0
\end{array}
\right)  ,
\]%
\[
\left(  M_{\dot{a}\dot{b}}^{-1}\right)  =\left(
\begin{array}
[c]{cc}%
0 & \delta_{s^{\prime}}^{s}\delta_{\tilde{b}}^{\tilde{a}}\\
-\delta_{s^{\prime}}^{s}\delta_{\tilde{b}}^{\tilde{a}} & 0
\end{array}
\right)  ,
\]
which shows that the coordinates are separated in two sets:%
\begin{align*}
t_{\dot{a}}  &  =\left\{  x_{s}^{\tilde{a}};\pi_{\tilde{a}}^{s}\right\}  ,\\
t_{\beta}  &  =\left\{  v^{a}\right\}  .
\end{align*}

The integrability conditions must be determined:%
\[
\left\{  \phi^{\beta},\phi^{t}\right\}  =\left\{  \phi^{\beta},\phi^{\dot{a}%
}\right\}  M_{\dot{a}\dot{b}}^{-1}\left\{  \phi^{\dot{b}},\phi^{t}\right\}  .
\]
Since $\left\{  \phi^{\dot{b}},\phi_{a}^{v}\right\}  =0$, it follows%
\begin{equation}
\left\{  \phi^{\beta},\phi^{t}\right\}  =0\Rightarrow\left\{  \phi_{b}%
^{v},\phi^{t}\right\}  =\frac{\partial H^{v}}{\partial v^{b}}=0\Rightarrow
\left\{
\begin{array}
[c]{c}%
\pi_{\tilde{a}}^{N_{\tilde{a}}}-\frac{\partial L^{v}}{\partial v^{\tilde{a}}%
}=0\\
\frac{\partial L^{v}}{\partial v^{\bar{a}}}=0
\end{array}
\right.  . \label{intcond2}%
\end{equation}
With these conditions some of the Lagrange multipliers are fixed.

Now the Generalized Brackets can be constructed, and for this, $\left\{
\phi^{\dot{a}},\phi^{t}\right\}  $ must be determined:%
\[
\left\{  \phi^{\dot{a}},\phi^{t}\right\}  \rightarrow\left\{
\begin{array}
[c]{c}%
\left\{  \phi_{\tilde{a}}^{x_{s}},\phi^{t}\right\}  =-\frac{\partial\phi^{t}%
}{\partial x_{s}^{\tilde{a}}}\\
\left\{  \phi_{\pi_{s}}^{\tilde{a}},\phi^{t}\right\}  =-\frac{\partial\phi
^{t}}{\partial\pi_{\tilde{a}}^{s}}%
\end{array}
\right.  .
\]
The total differential of $E=E\left(  x_{s}^{\tilde{a}},v^{a},\pi_{\tilde{a}%
}^{s}\right)  $ is%
\[
dE=\left\{  E,\phi^{\beta^{\prime}}\right\}  _{\ast}dt_{\beta^{\prime}%
}=\left\{  E,\phi^{t}\right\}  _{\ast}dt+\left\{  E,\phi_{a}^{v}\right\}
_{\ast}dv^{a},
\]
Since for any pair of functions $E=E\left(  x_{s}^{\tilde{a}},v^{a}%
,\pi_{\tilde{a}}^{s}\right)  $ and $F=F\left(  x_{s}^{\tilde{a}},v^{a}%
,\pi_{\tilde{a}}^{s}\right)  $\ one has $\left\{  E,F\right\}  =0$, it
follows, considering $H^{v}=H^{v}\left(  x_{s}^{\tilde{a}},v^{a},t;\pi
_{\tilde{a}}^{s}\right)  $,\ that%
\begin{equation}
dE=\frac{\partial E}{\partial t}dt+\frac{\partial E}{\partial v^{a}}%
dv^{a}+\left[  \frac{\partial E}{\partial x_{s}^{\tilde{c}}}\frac{\partial
\phi^{t}}{\partial\pi_{\tilde{c}}^{s}}-\frac{\partial\phi^{t}}{\partial
x_{s}^{\tilde{c}}}\frac{\partial E}{\partial\pi_{\tilde{c}}^{s}}\right]  dt.
\label{dE2}%
\end{equation}

At last the equations of motion are obtained:%
\[
dt_{\dot{a}}=\frac{\partial t_{\dot{a}}}{\partial t}dt+\frac{\partial
t_{\dot{a}}}{\partial v^{a}}dv^{a}+\left[  \frac{\partial t_{\dot{a}}%
}{\partial x_{s}^{\tilde{c}}}\frac{\partial\phi^{t}}{\partial\pi_{\tilde{c}%
}^{s}}-\frac{\partial\phi^{t}}{\partial x_{s}^{\tilde{c}}}\frac{\partial
t_{\dot{a}}}{\partial\pi_{\tilde{c}}^{s}}\right]  dt.
\]%
\begin{equation}
dx_{s}^{\tilde{a}}=\frac{\partial\phi^{t}}{\partial\pi_{\tilde{a}}^{s}%
}dt=\frac{\partial H^{v}}{\partial\pi_{\tilde{a}}^{s}}dt, \label{dx2}%
\end{equation}%
\begin{equation}
d\pi_{\tilde{a}}^{s}=-\frac{\partial\phi^{t}}{\partial x_{s}^{\tilde{a}}%
}dt=-\frac{\partial H^{v}}{\partial x_{s}^{\tilde{a}}}dt\Rightarrow\left\{
\begin{array}
[c]{l}%
d\pi_{\tilde{a}}^{s}=-\left(  \pi_{\tilde{a}}^{s-1}-\frac{\partial L^{v}%
}{\partial x_{s}^{\tilde{a}}}\right)  dt,\;\;s=2,...,N_{\tilde{a}},\\
d\pi_{\tilde{a}}^{1}=\frac{\partial L^{v}}{\partial x_{1}^{\tilde{a}}}dt.
\end{array}
\right.  \label{dpi2}%
\end{equation}
One observes that the remaining Lagrange multipliers (those not determined by
integrability conditions) are now fixed by the equations of motion:%
\begin{equation}
\pi_{\tilde{a}}^{s-1}=-\dot{\pi}_{\tilde{a}}^{s}+\frac{\partial L^{v}%
}{\partial x_{s}^{\tilde{a}}},\;\;s=2,...,N_{\tilde{a}}. \label{pi2}%
\end{equation}

\section{Examples}

\subsection{Podolsky Electrodynamics}

As it can be observed, when no degenerate coordinates exist, both approaches
coincide. As a first example it will be considered the case of Podolsky
Electrodynamics, which is a theory with second derivatives and no degenerate
coordinates. The intention is to show that the results obtained with the first
order approach is consistent with the results obtained by the standard HJ
approach for systems with higher order derivatives \cite{Rand Pim1} and those
obtained with Dirac's approach \cite{PimGalv}. Podolsky Electrodynamics
Lagrangian is given by:%
\begin{equation}
L=-\frac{1}{4}F_{\mu\nu}F^{\mu\nu}+a^{2}\partial_{\rho}F^{\mu\rho}%
\partial_{\sigma}F_{\mu}^{.\sigma}, \label{L0P}%
\end{equation}
where $F_{\mu\nu}=\partial_{\mu}A_{\nu}-\partial_{\nu}A_{\mu}$. $L$ can be
rewritten as
\begin{align}
L  &  =-\frac{1}{4}\left[  2\eta^{00}\eta^{ij}\left(  \partial_{0}%
A_{i}-\partial_{i}A_{0}\right)  \left(  \partial_{0}A_{j}-\partial_{j}%
A_{0}\right)  +F_{ij}F^{ij}\right]  +\nonumber\\
&  +a^{2}\left[  \eta^{ij}\eta^{00}\eta^{00}\left(  \partial_{j}\partial
_{0}A_{0}-\partial_{0}\partial_{0}A_{j}\right)  \left(  \partial_{i}%
\partial_{0}A_{0}-\partial_{0}\partial_{0}A_{i}\right)  \right.  +\nonumber\\
&  +\eta^{ik}\eta^{00}\eta^{jm}\left(  \partial_{i}\partial_{0}A_{k}%
-\partial_{i}\partial_{k}A_{0}\right)  \left(  \partial_{j}\partial_{0}%
A_{m}-\partial_{j}\partial_{m}A_{0}\right)  +\nonumber\\
&  +2\eta^{ik}\eta^{00}\eta^{jm}\left(  \partial_{k}\partial_{0}A_{0}%
-\partial_{0}\partial_{0}A_{k}\right)  \partial_{j}F_{im}+\nonumber\\
&  \left.  \left.  +\eta^{jl}\eta^{im}\eta^{kn}\partial_{i}F_{lm}\partial
_{k}F_{jn}\right)  \right]  , \label{L1P}%
\end{align}
which shows explicitly the time derivatives of $A_{\nu}$. It is immediate to
identify that $L=L\left(  A_{0},A_{i},\partial_{0}A_{0},\partial_{0}%
A_{i},\partial_{0}^{2}A_{i}\right)  $, \textit{i.e .}$N_{A_{0}}=1,N_{A_{i}}%
=2$. Introducing new variables%
\begin{align*}
x_{0}^{\left(  1\right)  }  &  \equiv A_{0},\;\;\;\;\;v_{0}\equiv\partial
_{0}A_{0}=\partial_{0}x_{0}^{\left(  1\right)  },\\
x_{i}^{\left(  1\right)  }  &  \equiv A_{i},\\
x_{i}^{\left(  2\right)  }  &  \equiv\partial_{0}A_{i}=\partial_{0}%
x_{i}^{\left(  1\right)  },\;\;\;\;\;v_{i}\equiv\partial_{0}\partial_{0}%
A_{i}=\partial_{0}x_{i}^{\left(  2\right)  },
\end{align*}
and defining%
\[
F_{ij}^{x_{1}}\equiv\partial_{i}x_{j}^{\left(  1\right)  }-\partial_{j}%
x_{i}^{\left(  1\right)  },
\]
one obtains%
\begin{align}
L^{v}  &  =L|_{A=x,\partial A=x,\partial^{2}A=v}=-\frac{1}{4}\left[
2\eta^{00}\eta^{ij}\left(  x_{i}^{\left(  2\right)  }-\partial_{i}%
x_{0}^{\left(  1\right)  }\right)  \left(  x_{j}^{\left(  2\right)  }%
-\partial_{j}x_{0}^{\left(  1\right)  }\right)  +F_{ij}^{x_{1}}F_{x_{1}}%
^{ij}\right]  +\nonumber\\
&  +a^{2}\left[  \eta^{ij}\eta^{00}\eta^{00}\left(  \partial_{j}v_{0}%
-v_{j}\right)  \left(  \partial_{i}v_{0}-v_{i}\right)  \right.  +\nonumber\\
&  +\eta^{ik}\eta^{00}\eta^{jm}\left(  \partial_{i}x_{k}^{\left(  2\right)
}-\partial_{i}\partial_{k}x_{0}^{\left(  1\right)  }\right)  \left(
\partial_{j}x_{m}^{\left(  2\right)  }-\partial_{j}\partial_{m}x_{0}^{\left(
1\right)  }\right)  +\nonumber\\
&  +2\eta^{ik}\eta^{00}\eta^{jm}\left(  \partial_{k}v_{0}-v_{k}\right)
\partial_{j}F_{im}^{x_{1}}+\nonumber\\
&  \left.  \left.  +\eta^{jl}\eta^{im}\eta^{kn}\partial_{i}F_{lm}^{x_{1}%
}\partial_{k}F_{jn}^{x_{1}}\right)  \right]  . \label{LvP}%
\end{align}

A new Lagrangian in an extended space can be constructed where Lagrange
multipliers are introduced%
\begin{align}
\bar{L}^{v}  &  =L^{v}+\pi_{\left(  1\right)  }^{k}\left(  \partial_{0}%
x_{k}^{\left(  1\right)  }-x_{k}^{\left(  2\right)  }\right)  +\pi_{\left(
2\right)  }^{k}\left(  \partial_{0}x_{k}^{\left(  2\right)  }-v_{k}\right)
+\pi_{\left(  1\right)  }^{0}\left(  \partial_{0}x_{0}^{\left(  1\right)
}-v_{0}\right) \nonumber\\
&  =\pi_{\left(  1\right)  }^{k}\partial_{0}x_{k}^{\left(  1\right)  }%
+\pi_{\left(  2\right)  }^{k}\partial_{0}x_{k}^{\left(  2\right)  }%
+\pi_{\left(  1\right)  }^{0}\partial_{0}x_{0}^{\left(  1\right)  }-H^{v},
\label{LvbarP}%
\end{align}%
\[
H^{v}=\pi_{\left(  1\right)  }^{k}x_{k}^{\left(  2\right)  }+\pi_{\left(
2\right)  }^{k}v_{k}+\pi_{\left(  1\right)  }^{0}v_{0}-L^{v},
\]
so that the configuration space now has "coordinates" $Z_{A}=\{x_{0}^{\left(
1\right)  },$ $x_{k}^{\left(  1\right)  },$ $x_{k}^{\left(  2\right)  },$
$v_{0},$ $v_{k},$ $\pi_{\left(  1\right)  }^{0},$ $\pi_{\left(  1\right)
}^{k},$ $\pi_{\left(  2\right)  }^{k}\}$, which are separated in the sets%
\begin{align*}
t_{b}  &  =\left\{  x_{0}^{\left(  1\right)  },x_{k}^{\left(  1\right)
},x_{k}^{\left(  2\right)  },\pi_{\left(  1\right)  }^{0},\pi_{\left(
1\right)  }^{k},\pi_{\left(  2\right)  }^{k}\right\}  ,\\
t_{\beta}  &  =\left\{  v_{0},v_{k}\right\}  ,
\end{align*}
with the following set of constraints%
\begin{align*}
\phi_{\left(  1\right)  }^{0}  &  =p_{\left(  1\right)  }^{0}-\pi_{\left(
1\right)  }^{0}=0,~\ \ \ \ \ \ \ \ \ \ \ \phi_{\left(  1\right)  }%
^{k}=p_{\left(  1\right)  }^{k}-\pi_{\left(  1\right)  }^{k}=0,\\
\phi_{\left(  2\right)  }^{k}  &  =p_{\left(  2\right)  }^{k}-\pi_{\left(
2\right)  }^{k}=0,~\ \ \ \ \ \ \ \ \ \ \ \phi_{v}^{0}=p_{v}^{0}=0,\\
\phi_{v}^{k}  &  =p_{v}^{k}=0,~\ \ \ \ \ \ \ \ \ \ \ \phi_{0}^{\pi_{\left(
1\right)  }}=p_{0}^{\pi_{\left(  1\right)  }}=0,\\
\phi_{k}^{\pi_{\left(  1\right)  }}  &  =p_{k}^{\pi_{\left(  1\right)  }%
}=0,~\ \ \ \ \ \ \ \ \ \ \ \phi_{k}^{\pi_{\left(  2\right)  }}=p_{k}%
^{\pi_{\left(  2\right)  }}=0,
\end{align*}%
\[
\phi^{t}=p^{0}+H^{v}=0.
\]

The integrability conditions must be evaluated. The first one gives%
\[
\frac{\delta H^{v}\left(  y\right)  }{\delta v_{0}\left(  x\right)
}=0\Rightarrow\pi_{\left(  1\right)  }^{0}=-2a^{2}\eta^{00}\eta^{ik}%
\partial_{k}\left[  \eta^{00}\left(  \partial_{i}v_{0}-v_{i}\right)
+\eta^{jm}\partial_{j}F_{im}^{x_{1}}\right]  .
\]
The other conditions are%
\begin{gather*}
\pi_{\left(  1\right)  }^{k}\delta_{k}^{n}\delta\left(  y-x\right)
-\frac{\delta L^{v}\left(  y\right)  }{\delta v_{n}\left(  x\right)
}=0\Rightarrow\\
\Rightarrow\pi_{\left(  1\right)  }^{n}=-2a^{2}\eta^{00}\eta^{ni}\left[
\eta^{00}\left(  \partial_{i}v_{0}-v_{i}\right)  +\eta^{jm}\partial_{j}%
F_{im}^{x_{1}}\right]  .
\end{gather*}

At this point one observation must be made. From this last expression one can
see that $v_{i}$\ can be written as a function of $\pi_{\left(  1\right)
}^{n}$,$v_{0}$\ and $F_{im}^{x_{1}}$:%
\[
v_{i}=\frac{1}{2a^{2}}\eta_{ni}\pi_{\left(  1\right)  }^{n}+\eta_{00}\eta
^{jm}\partial_{j}F_{im}^{x_{1}}+\partial_{i}v_{0}.
\]
The same cannot be said about $v_{0}$, which remains undetermined. When one
substitutes this result in $\pi_{\left(  1\right)  }^{0}$ it follows%
\[
\pi_{\left(  1\right)  }^{0}=\partial_{k}\pi_{\left(  1\right)  }^{k}.
\]

The Generalized Brackets can be evaluated when the following brackets are
calculated%
\begin{align*}
\left\{  \phi_{\left(  1\right)  }^{0}\left(  x\right)  ,\phi^{t}\left(
y\right)  \right\}   &  =-\int dz\frac{\delta H^{v}\left(  y\right)  }{\delta
x_{0}^{\left(  1\right)  }\left(  z\right)  }\delta\left(  z-x\right)  =\\
&  =\eta^{00}\eta^{ij}\left(  x_{j}^{\left(  2\right)  }-\partial_{j}%
x_{0}^{\left(  1\right)  }\right)  \partial_{i}^{y}\delta\left(  y-x\right)
+\\
&  -2a^{2}\eta^{00}\eta^{ik}\eta^{jm}\left(  \partial_{j}x_{m}^{\left(
2\right)  }-\partial_{j}\partial_{m}x_{0}^{\left(  1\right)  }\right)
\partial_{i}^{y}\partial_{k}^{y}\delta\left(  y-x\right)  ;
\end{align*}%
\begin{align*}
\left\{  \phi_{\left(  1\right)  }^{n}\left(  x\right)  ,\phi^{t}\left(
y\right)  \right\}   &  =-\eta^{ik}\eta^{jn}F_{ij}^{x_{1}}\partial_{k}%
^{y}\delta\left(  y-x\right)  +\\
&  -2a^{2}\left[  \eta^{nk}\eta^{ji}-\eta^{ik}\eta^{jn}\right]  \eta
^{00}\left(  \partial_{k}v_{0}-v_{k}\right)  \partial_{j}^{y}\partial_{i}%
^{y}\delta\left(  y-x\right)  +\\
&  -2a^{2}\left[  \eta^{jn}\eta^{im}-\eta^{jm}\eta^{in}\right]  \eta
^{kp}\partial_{k}F_{jp}^{x_{1}}\partial_{i}^{y}\partial_{m}^{y}\delta\left(
y-x\right)  ;
\end{align*}%
\begin{align*}
\left\{  \phi_{\left(  2\right)  }^{n}\left(  x\right)  ,\phi^{t}\left(
y\right)  \right\}   &  =-\pi_{\left(  1\right)  }^{n}\delta\left(
y-x\right)  -\eta^{00}\eta^{in}\left(  x_{i}^{\left(  2\right)  }-\partial
_{i}x_{0}^{\left(  1\right)  }\right)  \delta\left(  y-x\right)  +\\
&  +2a^{2}\eta^{ik}\eta^{00}\eta^{jn}\left(  \partial_{i}x_{k}^{\left(
2\right)  }-\partial_{i}\partial_{k}x_{0}^{\left(  1\right)  }\right)
\partial_{j}^{y}\delta\left(  y-x\right)  ;
\end{align*}%
\[
\left\{  \phi_{0}^{\pi_{\left(  1\right)  }}\left(  x\right)  ,\phi^{t}\left(
y\right)  \right\}  =-\frac{\delta H^{v}\left(  y\right)  }{\delta\pi_{\left(
1\right)  }^{0}\left(  x\right)  }=-v_{0}\delta\left(  y-x\right)  ;
\]%
\[
\left\{  \phi_{n}^{\pi_{\left(  1\right)  }}\left(  x\right)  ,\phi^{t}\left(
y\right)  \right\}  =-\frac{\delta H^{v}\left(  y\right)  }{\delta\pi_{\left(
1\right)  }^{n}\left(  x\right)  }=-x_{n}^{\left(  2\right)  }\delta\left(
y-x\right)  ;
\]%
\[
\left\{  \phi_{n}^{\pi_{\left(  2\right)  }}\left(  x\right)  ,\phi^{t}\left(
y\right)  \right\}  =-\frac{\delta H^{v}\left(  y\right)  }{\delta\pi_{\left(
2\right)  }^{n}\left(  x\right)  }=-v_{n}\delta\left(  y-x\right)  .
\]

The differential of any function $E$\ on this extended space is%
\begin{align*}
dE  &  =\frac{\partial E}{\partial t}dt+\left\{  E,H^{v}\right\}  dt+\left\{
E,\phi_{v}^{0}\right\}  dv_{0}+\left\{  E,\phi_{v}^{k}\right\}  dv_{k}+\\
&  +\int dw\left(  \left\{  E,\phi_{\left(  1\right)  }^{0}\left(  w\right)
\right\}  v_{0}+\left\{  E,\phi_{\left(  1\right)  }^{k}\left(  w\right)
\right\}  x_{k}^{\left(  2\right)  }+\left\{  E,\phi_{\left(  2\right)  }%
^{k}\left(  w\right)  \right\}  v_{k}+\right. \\
&  -\left\{  E,\phi_{0}^{\pi_{\left(  1\right)  }}\left(  w\right)  \right\}
\eta^{00}\eta^{ij}\partial_{i}^{w}\left(  x_{j}^{\left(  2\right)  }%
-\partial_{j}x_{0}^{\left(  1\right)  }\right)  +\\
&  -\left\{  E,\phi_{0}^{\pi_{\left(  1\right)  }}\left(  w\right)  \right\}
2a^{2}\eta^{00}\eta^{ik}\eta^{jm}\partial_{i}^{w}\partial_{k}^{w}\left(
\partial_{j}x_{m}^{\left(  2\right)  }-\partial_{j}\partial_{m}x_{0}^{\left(
1\right)  }\right)  +\\
&  +\left\{  E,\phi_{n}^{\pi_{\left(  1\right)  }}\left(  w\right)  \right\}
\eta^{ik}\eta^{jn}\partial_{k}^{w}F_{ij}^{x_{1}}+\\
&  -\left\{  E,\phi_{n}^{\pi_{\left(  1\right)  }}\left(  y\right)  \right\}
2a^{2}\left[  \eta^{nk}\eta^{ji}-\eta^{ik}\eta^{jn}\right]  \eta^{00}%
\partial_{j}^{w}\partial_{i}^{w}\left(  \partial_{k}v_{0}-v_{k}\right)  +\\
&  -\left\{  E,\phi_{n}^{\pi_{\left(  1\right)  }}\left(  w\right)  \right\}
2a^{2}\left[  \eta^{jn}\eta^{im}-\eta^{jm}\eta^{in}\right]  \eta^{kp}%
\partial_{i}^{w}\partial_{m}^{w}\partial_{k}F_{jp}^{x_{1}}+\\
&  -\left\{  E,\phi_{n}^{\pi_{\left(  2\right)  }}\left(  w\right)  \right\}
\pi_{\left(  1\right)  }^{n}-\left\{  E,\phi_{n}^{\pi_{\left(  2\right)  }%
}\left(  w\right)  \right\}  \eta^{00}\eta^{in}\left(  x_{i}^{\left(
2\right)  }-\partial_{i}x_{0}^{\left(  1\right)  }\right)  +\\
&  \left.  -\left\{  E,\phi_{n}^{\pi_{\left(  2\right)  }}\left(  w\right)
\right\}  2a^{2}\eta^{ik}\eta^{00}\eta^{jn}\partial_{j}^{w}\left(
\partial_{i}x_{k}^{\left(  2\right)  }-\partial_{i}\partial_{k}x_{0}^{\left(
1\right)  }\right)  \right)  dt\left(  w\right)  .
\end{align*}

To obtain the equations of motion, it is necessary to consider $E=\{x_{0}%
^{\left(  1\right)  },$ $x_{k}^{\left(  1\right)  },$ $x_{k}^{\left(
2\right)  },$ $\pi_{\left(  1\right)  }^{0},$ $\pi_{\left(  1\right)  }^{k},$
$\pi_{\left(  2\right)  }^{k}\}$:%
\begin{equation}
dx_{0}^{\left(  1\right)  }=v_{0}\left(  z\right)  dt; \label{dx01P}%
\end{equation}%
\begin{equation}
dx_{p}^{\left(  1\right)  }=x_{p}^{\left(  2\right)  }dt; \label{dxk2P}%
\end{equation}%
\begin{equation}
dx_{p}^{\left(  2\right)  }=v_{p}dt=\left[  \frac{1}{2a^{2}}\eta_{np}%
\pi_{\left(  1\right)  }^{n}+\eta_{00}\eta^{jm}\partial_{j}F_{pm}^{x_{1}%
}+\partial_{p}v_{0}\right]  dt; \label{dxp2P}%
\end{equation}
\bigskip%
\begin{equation}
d\pi_{\left(  1\right)  }^{0}=-\left[  \eta^{00}\eta^{ij}\partial_{i}\left(
x_{j}^{\left(  2\right)  }-\partial_{j}x_{0}^{\left(  1\right)  }\right)
+2a^{2}\eta^{00}\eta^{ik}\eta^{jm}\partial_{i}\partial_{k}\left(  \partial
_{j}x_{m}^{\left(  2\right)  }-\partial_{j}\partial_{m}x_{0}^{\left(
1\right)  }\right)  \right]  dt; \label{dpi01P}%
\end{equation}%
\begin{equation}
d\pi_{\left(  1\right)  }^{p}=\left(  \eta^{ik}\eta^{jp}\partial_{k}%
F_{ij}^{x_{1}}+\eta_{00}\left[  \eta^{ij}\partial_{i}\partial_{j}\pi_{\left(
1\right)  }^{p}-\eta^{ip}\partial_{i}\pi_{\left(  1\right)  }^{0}\right]
\right)  dt; \label{dpip1P}%
\end{equation}%
\begin{align*}
d\pi_{\left(  2\right)  }^{p}  &  =-\left(  \pi_{\left(  1\right)  }^{p}%
+\eta^{00}\eta^{ip}\left(  x_{i}^{\left(  2\right)  }-\partial_{i}%
x_{0}^{\left(  1\right)  }\right)  +\right. \\
&  \left.  +2a^{2}\eta^{ik}\eta^{00}\eta^{jp}\partial_{j}\left(  \partial
_{i}x_{k}^{\left(  2\right)  }-\partial_{i}\partial_{k}x_{0}^{\left(
1\right)  }\right)  \right)  dt.
\end{align*}

All the results obtained here are in accordance with those obtained in
\cite{Rand Pim1} and \cite{PimGalv}, showing the consistency of the
construction made in this work with the non first order HJ approach and with
Dirac's procedure.

\subsection{Proca Model}

The next examples will be used to compare the two approaches introduced in
this work.

Let us now consider the case of Proca Model, whose Lagrangian is%
\[
L=-\frac{1}{4}F_{\mu\nu}F^{\mu\nu}+\frac{f}{2}m^{2}A_{\mu}A^{\mu},
\]%
\begin{align}
L  &  =-\frac{1}{4}\left[  2\eta^{00}\eta^{ij}\left(  \partial_{0}%
A_{i}-\partial_{i}A_{0}\right)  \left(  \partial_{0}A_{j}-\partial_{j}%
A_{0}\right)  +\eta^{ik}\eta^{jn}F_{ij}F_{kn}\right]  +\nonumber\\
&  +\eta^{00}\frac{f}{2}m^{2}A_{0}A_{0}+\eta^{ij}\frac{f}{2}m^{2}A_{i}A_{j}
\label{LProca}%
\end{align}
where $f=\pm1$ according the convention of the metric, and $F_{\mu\nu
}=\partial_{\mu}A_{\nu}-\partial_{\nu}A_{\mu}$.

\subsubsection{Standard approach}

Introducing new variables%
\[
x_{\mu}^{\left(  1\right)  }\equiv A_{\mu},\;\;\;\;\;v_{\mu}\equiv\partial
_{0}A_{\mu}=\partial_{0}x_{\mu}^{\left(  1\right)  },
\]
and defining%
\[
F_{ij}^{x_{1}}\equiv\partial_{i}x_{j}^{\left(  1\right)  }-\partial_{j}%
x_{i}^{\left(  1\right)  },
\]
one finds%
\begin{align}
L^{v}  &  =-\frac{1}{4}\left[  2\eta^{00}\eta^{ij}\left(  v_{i}-\partial
_{i}x_{0}^{\left(  1\right)  }\right)  \left(  v_{j}-\partial_{j}%
x_{0}^{\left(  1\right)  }\right)  +\eta^{ik}\eta^{jn}F_{ij}^{x_{1}}%
F_{kn}^{x_{1}}\right]  +\nonumber\\
&  +\eta^{\mu\nu}\frac{f}{2}m^{2}x_{\mu}^{\left(  1\right)  }x_{\nu}^{\left(
1\right)  }. \label{LvPr}%
\end{align}

Now one can define a new Lagrangian $\bar{L}^{v}$ in an extended space with
"coordinates" $z_{A}=\left\{  x_{\mu}^{\left(  1\right)  },v_{\mu},\pi^{\mu
}\right\}  $ such that%
\begin{equation}
\bar{L}^{v}=L^{v}+\pi^{\mu}\left(  \partial_{0}x_{\mu}^{\left(  1\right)
}-v_{\mu}\right)  =\pi^{\mu}\partial_{0}x_{\mu}^{\left(  1\right)  }-H^{v},
\label{LvbarPr}%
\end{equation}%
\begin{equation}
H^{v}\equiv\pi^{\mu}v_{\mu}-L^{v}. \label{HvPr}%
\end{equation}
The constraints can be obtained:%
\begin{align*}
\phi_{x}^{\mu}  &  =p_{x}^{\mu}-\pi^{\mu}=0,~\ \ \ \ \ \ \ \ \ \ \phi_{v}%
^{\mu}=p_{v}^{\mu}=0,\\
\phi_{\mu}^{\pi}  &  =p_{\mu}^{\pi}=0,
\end{align*}%
\[
\phi^{t}=p^{0}+H^{v}=0.
\]

The variables then are separated in two sets%
\begin{align*}
t_{b}  &  =\left\{  x_{\mu}^{\left(  1\right)  };\pi^{\mu}\right\}  ,\\
t_{\beta}  &  =\left\{  v_{\mu}\right\}  ,
\end{align*}
and one must obtain the integrability conditions:%
\[
\pi^{\mu}\delta_{\mu}^{\sigma}\delta\left(  y-x\right)  -\frac{\delta
L^{v}\left(  y\right)  }{\delta v_{\sigma}\left(  x\right)  }=0,
\]%
\begin{equation}
\pi^{0}\left(  x\right)  =0 \label{cond1aPr}%
\end{equation}

\begin{equation}
\pi^{k}\left(  x\right)  =-\eta^{00}\eta^{ik}\left(  v_{i}\left(  x\right)
-\partial_{i}x_{0}^{\left(  1\right)  }\left(  x\right)  \right)
\label{cond2Pr}%
\end{equation}
From this last expression it follows%
\begin{equation}
v_{i}=\partial_{i}x_{0}^{\left(  1\right)  }-\eta_{00}\eta_{ki}\pi^{k},
\label{vi1Pr}%
\end{equation}
while $v_{0}$\ is not determined as a function of the other variables.

The total differential of any function $E=E\left(  z_{A}\right)  $ is given by%
\begin{align}
dE  &  =\left\{  E,\phi^{t}\right\}  dt+\left\{  E,\phi_{v}^{\mu}\right\}
dv_{\mu}+\nonumber\\
&  -\int dz\int dx\int dw\left(
\begin{array}
[c]{cc}%
\left\{  E,\phi_{x}^{\nu}\left(  z\right)  \right\}  & \left\{  E,\phi_{\nu
}^{\pi}\left(  z\right)  \right\}
\end{array}
\right)  \delta\left(  z-x\right)  .\nonumber\\
&  .\left(
\begin{array}
[c]{c}%
-v_{\nu}\left(  w\right)  \delta\left(  w-x\right) \\
\eta^{mn}\eta^{l\nu}F_{ml}^{x_{1}}\left(  w\right)  \partial_{n}^{w}%
\delta\left(  w-x\right)  -fm^{2}\eta^{\rho\nu}x_{\rho}^{\left(  1\right)
}\left(  w\right)  \delta\left(  w-x\right)  -C^{\nu}%
\end{array}
\right)  dt\left(  w\right)  , \label{dEPr}%
\end{align}
where%
\[
C^{\sigma}\equiv\eta^{00}\eta^{nm}\left(  v_{n}-\partial_{n}x_{0}^{\left(
1\right)  }\right)  \delta_{0}^{\sigma}\partial_{m}^{w}\delta\left(
w-x\right)  .
\]
The equations of motion can be evaluated:%
\begin{equation}
dx_{i}^{\left(  1\right)  }=\left(  \partial_{i}x_{0}^{\left(  1\right)
}-\eta_{00}\eta_{ki}\pi^{k}\right)  dt; \label{dxiPr}%
\end{equation}%
\begin{equation}
dx_{0}^{\left(  1\right)  }=v_{0}dt; \label{dx0Pr}%
\end{equation}%
\begin{equation}
d\pi^{i}=\left[  \eta^{mn}\eta^{li}\partial_{n}F_{ml}^{x_{1}}+fm^{2}\eta
^{mi}x_{m}^{\left(  1\right)  }\right]  dt \label{dpiiPr}%
\end{equation}%
\begin{equation}
d\pi^{0}=\left[  fm^{2}\eta^{00}x_{0}^{\left(  1\right)  }-\eta^{00}\eta
^{nm}\partial_{m}\left(  v_{n}-\partial_{n}x_{0}^{\left(  1\right)  }\right)
\right]  dt \label{dpi0Pr}%
\end{equation}

With (\ref{cond2Pr}) and (\ref{cond1aPr}) it follows%
\begin{align*}
d\pi^{0}  &  =\left[  fm^{2}\eta^{00}x_{0}^{\left(  1\right)  }+\partial
_{m}\pi^{m}\right]  dt=0\Rightarrow\\
&  \Rightarrow x_{0}^{\left(  1\right)  }=-\frac{\eta_{00}}{fm^{2}}%
\partial_{m}\pi^{m},
\end{align*}
so that $x_{0}^{\left(  1\right)  }$\ is written in terms of the $\pi^{m}$ and
thereafter%
\[
v_{i}=-\eta_{00}\left(  \frac{1}{fm^{2}}\partial_{i}\partial_{m}+\eta
_{mi}\right)  \pi^{m},
\]%
\[
dx_{i}^{\left(  1\right)  }=-\eta_{00}\left[  \frac{1}{fm^{2}}\partial
_{i}\partial_{m}\pi^{m}+\eta_{mi}\pi^{m}\right]  dt;
\]%
\[
d\pi^{i}=\left[  \eta^{mn}\eta^{li}\partial_{n}F_{ml}^{x_{1}}+fm^{2}\eta
^{mi}x_{m}^{\left(  1\right)  }\right]  dt.
\]

\subsubsection{Degenerate Coordinates Approach}

The Lagrangian (\ref{LProca}) shows that $A_{0}$ is a degenerate coordinate,
\textit{i.e.} $L=L\left(  A_{0},A_{i},\partial_{0}A_{i}\right)  $, and
$N_{A_{0}}=0,N_{A_{i}}=1$. Introducing new variables%
\begin{align*}
v_{0}  &  \equiv A_{0},\\
x_{i}^{\left(  1\right)  }  &  \equiv A_{i},\;\;\;\;\;v_{i}\equiv\partial
_{0}A_{i}=\partial_{0}x_{i}^{\left(  1\right)  },
\end{align*}
and defining%
\[
F_{ij}^{x_{1}}\equiv\partial_{i}x_{j}^{\left(  1\right)  }-\partial_{j}%
x_{i}^{\left(  1\right)  },
\]
it follows%
\begin{align}
L^{v}  &  =-\frac{1}{4}\left[  2\eta^{00}\eta^{ij}\left(  v_{i}-\partial
_{i}v_{0}\right)  \left(  v_{j}-\partial_{j}v_{0}\right)  +\eta^{ik}\eta
^{jn}F_{ij}^{x_{1}}F_{kn}^{x_{1}}\right]  +\nonumber\\
&  +\eta^{00}\frac{f}{2}m^{2}v_{0}v_{0}+\eta^{ij}\frac{f}{2}m^{2}%
x_{i}^{\left(  1\right)  }x_{j}^{\left(  1\right)  }. \label{LPr2}%
\end{align}

Now the new Lagrangian $\bar{L}^{v}$\ with Lagrange multipliers can be written
down:%
\begin{equation}
\bar{L}^{v}=L^{v}+\pi^{i}\left(  \partial_{0}x_{i}^{\left(  1\right)  }%
-v_{i}\right)  =\pi^{i}\partial_{0}x_{i}^{\left(  1\right)  }-H^{v},
\label{LvbarPr2a}%
\end{equation}%
\begin{equation}
H^{v}\equiv\pi^{i}v_{i}-L^{v}. \label{HvPr2}%
\end{equation}

The constraints are identified:%
\begin{align*}
\phi_{x}^{i}  &  =p_{x}^{i}-\pi^{i}=0,~\ \ \ \ \ \ \ \ \phi_{v}^{0}=p_{v}%
^{0}=0,\\
\phi_{v}^{i}  &  =p_{v}^{i}=0,~\ \ \ \ \ \ \ \ \phi_{i}^{\pi}=p_{i}^{\pi}=0,
\end{align*}%
\[
\phi^{t}=p^{0}+H^{v}=0,
\]
and the variables are separated:%
\begin{align*}
t_{b}  &  =\left\{  x_{i}^{\left(  1\right)  };\pi^{i}\right\}  ,\\
t_{\beta}  &  =\left\{  v_{0};v_{i}\right\}  .
\end{align*}

The integrability conditions must be determined:%
\begin{equation}
\frac{\delta L^{v}\left(  y\right)  }{\delta v_{0}\left(  x\right)  }%
=\eta^{00}\eta^{ij}\left(  v_{i}-\partial_{i}v_{0}\right)  \partial_{j}%
^{y}\delta\left(  y-x\right)  +\eta^{00}fm^{2}v_{0}\delta\left(  y-x\right)
=0, \label{cond1aPr2}%
\end{equation}%
\begin{equation}
\pi^{k}\left(  x\right)  =-\eta^{00}\eta^{ik}\left(  v_{i}\left(  x\right)
-\partial_{i}v_{0}\left(  x\right)  \right)  , \label{cond2aPr2}%
\end{equation}
from where it is immediate to verfiy that%
\begin{equation}
v_{i}=\partial_{i}v_{0}-\eta_{00}\eta_{ki}\pi^{k}. \label{viPr2}%
\end{equation}
Integrating (\ref{cond1aPr2}), one finds%
\begin{equation}
v_{0}=-\frac{\eta_{00}}{fm^{2}}\partial_{j}\pi^{j}, \label{v0Pr2}%
\end{equation}
which leads one to conclude that%
\begin{equation}
v_{i}=-\eta_{00}\left(  \frac{1}{fm^{2}}\partial_{i}\partial_{j}+\eta
_{ji}\right)  \pi^{j}. \label{vi2Pr2}%
\end{equation}

The differential of any function $E=E\left(  z_{A}\right)  $\ is%
\begin{align}
dE  &  =\left\{  E,\phi^{t}\right\}  dt+\left\{  E,\phi_{v}^{0}\right\}
dv_{0}+\left\{  E,\phi_{v}^{i}\right\}  dv_{i}+\nonumber\\
&  -\int dz\int dx\int dw\left(
\begin{array}
[c]{cc}%
\left\{  E,\phi_{x}^{j}\left(  z\right)  \right\}  & \left\{  E,\phi_{j}^{\pi
}\left(  z\right)  \right\}
\end{array}
\right)  \delta\left(  z-x\right)  .\nonumber\\
&  .\left(
\begin{array}
[c]{c}%
-v_{j}\left(  w\right)  \delta\left(  w-x\right) \\
\eta^{mn}\eta^{lj}F_{ml}^{x_{1}}\left(  w\right)  \partial_{n}^{w}%
\delta\left(  w-x\right)  -fm^{2}\eta^{mj}x_{m}^{\left(  1\right)  }\left(
w\right)  \delta\left(  w-x\right)
\end{array}
\right)  dt\left(  w\right)  , \label{dEPr2}%
\end{align}
and the equations of motion are:%
\begin{equation}
dx_{i}^{\left(  1\right)  }=-\eta_{00}\left(  \frac{1}{fm^{2}}\partial
_{i}\partial_{j}^{x}+\eta_{ji}\right)  \pi^{j}dt; \label{dxiPr2}%
\end{equation}%
\begin{equation}
d\pi^{i}=\left[  \eta^{mn}\eta^{li}\partial_{n}F_{ml}^{x_{1}}+fm^{2}\eta
^{mi}x_{m}^{\left(  1\right)  }\right]  dt. \label{dpiiPr2}%
\end{equation}

It is interesting to notice that in the standard approach, the determination
of $x_{0}^{\left(  1\right)  }$ (\textit{i.e. }$A_{0}$) and $v_{i}$ does not
occur when integrability conditions are evaluated, but it can only be obtained
when equations of motion are considered and when a kind of consistency
condition is applied ($\pi^{0}=0\Rightarrow d\pi^{0}=0$). In the degenerate
approach, $v_{0}$ (\textit{i.e. }$A_{0}$) and $v_{i}$ are readily determined
by integrability conditions, no use of equations of motion are necessary and
no extra condition must be applied.

\subsection{QCD}

The Lagrangian of the gauge field of QCD is%
\[
L=-\frac{1}{4}F_{\mu\nu}^{a}F_{a}^{\mu\nu},
\]
where $F_{\mu\nu}^{a}=\partial_{\mu}A_{\nu}^{a}-\partial_{\nu}A_{\mu}%
^{a}+f_{b\;c}^{\;a}A_{\mu}^{b}A_{\nu}^{c}$. Expliciting the time derivatives
one finds:%
\begin{align}
L  &  =-\frac{1}{2}\eta^{00}\eta^{ij}\left(  \partial_{0}A_{i}^{a}%
-\partial_{i}A_{0}^{a}+f_{b\;c}^{\;a}A_{0}^{b}A_{i}^{c}\right)  \partial
_{0}A_{aj}+\nonumber\\
&  -\frac{1}{2}\eta^{00}\eta^{ij}\left(  \partial_{0}A_{i}^{a}-\partial
_{i}A_{0}^{a}+f_{b\;c}^{\;a}A_{0}^{b}A_{i}^{c}\right)  \left(  -\partial
_{j}A_{a0}+f_{dae}A_{0}^{d}A_{j}^{e}\right)  +F_{ij}^{a}F_{a}^{ij}.
\label{LQCD}%
\end{align}

\subsubsection{Standard Approach}

With the new variables%
\[
x_{\mu}^{a}\equiv A_{\mu}^{a},\;\;\;\;\;v_{\mu}^{a}\equiv\partial_{0}A_{\mu
}^{a}=\partial_{0}x_{\mu}^{a},
\]
and with the definition%
\[
F_{ij}^{a\left(  x\right)  }\equiv\partial_{i}x_{j}^{a}-\partial_{j}x_{i}%
^{a}+f_{d~e}^{~a}x_{i}^{d}x_{j}^{e},
\]
it follows%
\begin{align}
L^{v}  &  =-\frac{1}{4}\left[  2\eta^{00}\eta^{ij}\left(  v_{i}^{a}%
-\partial_{i}x_{0}^{a}+f_{b\;c}^{\;a}x_{0}^{b}x_{i}^{c}\right)  v_{aj}+\right.
\nonumber\\
&  \left.  +2\eta^{00}\eta^{ij}\left(  v_{i}^{a}-\partial_{i}x_{0}%
^{a}+f_{b\;c}^{\;a}x_{0}^{b}x_{i}^{c}\right)  \left(  -\partial_{j}%
x_{a0}+f_{dae}x_{0}^{d}x_{j}^{e}\right)  +F_{ij}^{a\left(  x\right)
}F_{a\left(  x\right)  }^{ij}\right]  . \label{LvQ}%
\end{align}

\begin{equation}
\bar{L}^{v}=\pi_{a}^{\mu}\partial_{0}x_{\mu}^{a}-H^{v}, \label{LvbarQ}%
\end{equation}%
\begin{equation}
H^{v}\equiv\pi_{a}^{\mu}v_{\mu}^{a}-L^{v}. \label{HvQ}%
\end{equation}

The constraints are identified%
\begin{align*}
\phi_{\left(  x\right)  a}^{\mu}  &  =p_{\left(  x\right)  a}^{\mu}-\pi
_{a}^{\mu}=0,~\ \ \ \ \ \ \ \ \phi_{\left(  v\right)  a}^{\mu}=p_{\left(
v\right)  a}^{\mu}=0,\\
\phi_{\mu}^{\left(  \pi\right)  a}  &  =p_{\mu}^{\left(  \pi\right)  a}=0,
\end{align*}%
\[
\phi^{t}=p^{0}+H^{v}=0,
\]
and the variables are separated%
\begin{align*}
t_{b}  &  =\left\{  x_{\mu}^{a};\pi_{a}^{\mu}\right\}  ,\\
t_{\beta}  &  =\left\{  v_{\mu}^{a}\right\}  .
\end{align*}

The conditions that fix the subspace are%
\begin{equation}
\pi_{c}^{0}\delta\left(  y-x\right)  =\frac{\delta L^{v}\left(  y\right)
}{\delta v_{0}^{c}\left(  x\right)  }=0, \label{cond1aQ}%
\end{equation}%
\begin{equation}
\pi_{c}^{k}\left(  x\right)  =-\eta^{00}\eta^{ik}\left(  v_{ci}-\partial
_{i}x_{c0}+f_{bcd}x_{0}^{b}x_{i}^{d}\right)  , \label{cond2Q}%
\end{equation}
from where one obtains
\begin{equation}
v_{ci}=\partial_{i}x_{c0}-f_{bcd}x_{0}^{b}x_{i}^{d}-\eta_{00}\eta_{ik}\pi
_{c}^{k}, \label{vciQ}%
\end{equation}
while $v_{0}^{c}$\ remains undetermined.

$dE$ is given by%
\begin{align}
dE  &  =\left\{  E,\phi^{t}\right\}  dt+\left\{  E,\phi_{\left(  v\right)
a}^{\mu}\right\}  dv_{\mu}^{a}+\nonumber\\
&  -\int dz\int dy\int dw\left(
\begin{array}
[c]{cc}%
\left\{  E,\phi_{\left(  x\right)  a}^{\mu}\left(  z\right)  \right\}  &
\left\{  E,\phi_{\mu}^{\left(  \pi\right)  a}\left(  z\right)  \right\}
\end{array}
\right)  .\nonumber\\
&  .\left(
\begin{array}
[c]{c}%
-v_{\mu}^{a}\left(  w\right)  \delta\left(  w-y\right) \\
-\left(  C_{a}^{\mu}\left(  w-y\right)  +B_{a}^{\mu}\left(  w-y\right)
\right)
\end{array}
\right)  \delta\left(  z-y\right)  dt\left(  w\right)  , \label{dEQ}%
\end{align}
where%
\begin{align*}
C_{a}^{\mu}  &  \equiv-\eta^{00}\eta^{ij}\left(  v_{ei}-\partial_{i}^{w}%
x_{e0}+f_{def}x_{0}^{d}x_{i}^{f}\right)  .\\
&  .\left(  -\delta_{a}^{e}\delta_{0}^{\mu}\partial_{j}^{w}\delta\left(
w-y\right)  +f_{a\;c}^{\;e}\delta_{0}^{\mu}\delta\left(  w-y\right)  x_{j}%
^{c}+f_{b\;a}^{\;e}x_{0}^{b}\delta_{j}^{\mu}\delta\left(  w-y\right)  \right)
,\\
B_{a}^{\mu}  &  \equiv-\frac{1}{2}F_{e\left(  x\right)  }^{ij}\left(
\delta_{a}^{e}\delta_{j}^{\mu}\partial_{i}^{w}\delta\left(  w-y\right)
-\delta_{a}^{e}\delta_{i}^{\mu}\partial_{j}^{w}\delta\left(  w-y\right)
\right.  +\\
&  +\left.  f_{a~c}^{~e}\delta_{i}^{\mu}\delta\left(  w-y\right)  x_{j}%
^{c}+f_{d~a}^{~e}x_{i}^{d}\delta_{j}^{\mu}\delta\left(  w-y\right)  \right)  .
\end{align*}

The equations of motion are obtained:%
\[
dx_{\rho}^{g}\left(  u\right)  =v_{\rho}^{g}\left(  u\right)  dt,
\]%
\begin{equation}
dx_{i}^{g}=\left(  \partial_{i}x_{0}^{g}-f_{b~d}^{~g}x_{0}^{b}x_{i}^{d}%
-\eta_{00}\eta_{ik}\pi^{gk}\right)  dt, \label{dxiQ}%
\end{equation}%
\begin{equation}
dx_{0}^{g}=v_{0}^{g}dt; \label{dx0Q}%
\end{equation}

\begin{align*}
d\pi_{a}^{\mu}\left(  u\right)   &  =-\eta^{00}\eta^{ij}\left[  \delta
_{0}^{\mu}D_{ja}^{e}\left(  v_{ei}-\partial_{i}x_{e0}+f_{def}x_{0}^{d}%
x_{i}^{f}\right)  \right.  +\\
&  +\left.  \delta_{j}^{\mu}f_{b\;a}^{\;e}x_{0}^{b}\left(  v_{ei}-\partial
_{i}x_{e0}+f_{def}x_{0}^{d}x_{i}^{f}\right)  \right]  \left(  u\right)
dt\left(  u\right)  +\\
&  +\delta_{j}^{\mu}D_{ia}^{e}F_{e\left(  x\right)  }^{ij}\left(  u\right)
dt\left(  u\right)  ,
\end{align*}
where%
\[
D_{ia}^{e}F_{e\left(  x\right)  }^{ij}=\delta_{a}^{e}\partial_{i}F_{e\left(
x\right)  }^{ij}+f_{a~c}^{~e}x_{i}^{c}F_{e\left(  x\right)  }^{ij},
\]
which leads to%
\begin{equation}
d\pi_{a}^{0}=-\eta^{00}\eta^{ij}D_{ja}^{e}\left(  v_{ei}-\partial_{i}%
^{w}x_{e0}+f_{def}x_{0}^{d}x_{i}^{f}\right)  dt, \label{dpi0Q}%
\end{equation}%
\begin{equation}
d\pi_{a}^{k}\left(  u\right)  =-\eta^{00}\eta^{ik}f_{b\;a}^{\;e}x_{0}%
^{b}\left(  v_{ei}-\partial_{i}^{w}x_{e0}+f_{def}x_{0}^{d}x_{i}^{f}\right)
\left(  u\right)  dt+D_{ia}^{e}F_{e\left(  x\right)  }^{ik}dt. \label{dpikQ}%
\end{equation}

With conditions (\ref{cond2Q}) and (\ref{cond1aQ}) it follows%
\[
d\pi_{a}^{k}\left(  u\right)  =f_{b\;a}^{\;e}x_{0}^{b}\pi_{e}^{k}dt+D_{ia}%
^{e}F_{e\left(  x\right)  }^{ik}dt,
\]%
\[
d\pi_{a}^{0}=D_{ja}^{e}\pi_{e}^{j}dt=0\Rightarrow D_{ja}^{e}\pi_{e}^{j}=0.
\]
This last result is the non-abelian generalization of the Gauss law and it
arises only when one considers $d\pi_{a}^{0}=0$.

\subsubsection{Degenerate Coordinates Approach}

Since $L=L\left(  A_{0}^{a},A_{i}^{a},\partial_{0}A_{i}^{a}\right)  $,
\textit{i.e.} $N_{A_{0}^{a}}=0,N_{A_{i}^{a}}=1$, the new variables will be%
\begin{align*}
v_{0}^{a}  &  \equiv A_{0}^{a},\\
x_{i}^{a}  &  \equiv A_{i}^{a},\;\;\;\;\;v_{i}^{a}\equiv\partial_{0}A_{i}%
^{a}=\partial_{0}x_{i}^{a}.
\end{align*}
With%
\[
F_{ij}^{a\left(  x\right)  }\equiv\partial_{i}x_{j}^{a}-\partial_{j}x_{i}%
^{a}+f_{dae}x_{i}^{d}x_{j}^{e},
\]
it follows%
\begin{align}
L^{v}  &  =-\frac{1}{4}\left[  2\eta^{00}\eta^{ij}v_{i}^{a}v_{aj}+4\eta
^{00}\eta^{ij}v_{i}^{a}\left(  -\partial_{j}v_{a0}+f_{dae}v_{0}^{d}x_{j}%
^{e}\right)  +\right. \nonumber\\
&  \left.  +2\eta^{00}\eta^{ij}\left[  -\partial_{i}v_{0}^{a}+f_{b\;c}%
^{\;a}v_{0}^{b}x_{i}^{c}\right]  \left[  -\partial_{j}v_{a0}+f_{dae}v_{0}%
^{d}x_{j}^{e}\right]  +F_{ij}^{a\left(  x\right)  }F_{a\left(  x\right)
}^{ij}\right]  , \label{LvQ2}%
\end{align}

\begin{equation}
\bar{L}^{v}=\pi_{a}^{i}\partial_{0}x_{i}^{a}-H^{v}, \label{LvbarQ2}%
\end{equation}%
\begin{equation}
H^{v}\equiv\pi_{a}^{i}v_{i}^{a}-L^{v}, \label{HvQ2}%
\end{equation}
and%
\begin{align*}
\phi_{a\left(  x\right)  }^{i}  &  =p_{a\left(  x\right)  }^{i}-\pi_{a}%
^{i}=0,~\ \ \ \ \ \ \ \ \ \phi_{a\left(  v\right)  }^{0}=p_{a\left(  v\right)
}^{0}=0,\\
\phi_{a\left(  v\right)  }^{i}  &  =p_{a\left(  v\right)  }^{i}%
=0,~\ \ \ \ \ \ \ \ \ \phi_{i}^{a\left(  \pi\right)  }=p_{i}^{a\left(
\pi\right)  }=0,
\end{align*}%
\[
\phi^{t}=p^{0}+H^{v}=0.
\]

One has%
\begin{align*}
t_{b_{x}}  &  =\left\{  x_{i}^{a};\pi_{a}^{i}\right\}  ,\\
t_{\beta_{x}}  &  =\left\{  v_{0}^{a};v_{i}^{a}\right\}  .
\end{align*}

The integrability conditions are%
\begin{gather}
\eta^{00}\eta^{ij}\partial_{j}\partial_{i}v_{g0}-\eta^{00}\eta^{ij}%
\partial_{j}v_{gi}-\eta^{00}\eta^{ij}f_{gae}v_{i}^{a}x_{j}^{e}+\eta^{00}%
\eta^{ij}f_{gac}x_{j}^{c}\partial_{i}v_{0}^{a}+\nonumber\\
-\eta^{00}\eta^{ij}f_{agc}x_{j}^{c}\partial_{i}v_{0}^{a}-\eta^{00}\eta
^{ij}f_{agc}v_{0}^{a}\partial_{i}x_{j}^{c}-\eta^{00}\eta^{ij}f_{g\;c}%
^{\;a}f_{bae}v_{0}^{b}x_{j}^{e}x_{i}^{c}=0, \label{cond1Q2}%
\end{gather}%
\begin{equation}
\pi_{g}^{k}\left(  x\right)  =-\eta^{00}v_{g}^{k}+\eta^{00}\partial^{k}%
v_{g0}-\eta^{00}f_{dge}v_{0}^{d}x^{ek}. \label{cond2Q2}%
\end{equation}
From this last result one finds
\begin{equation}
v_{g}^{k}=-\eta_{00}\pi_{g}^{k}+\partial^{k}v_{g0}-f_{dge}v_{0}^{d}x^{ek},
\label{vi1Q2}%
\end{equation}
which, with the previous condition, leads to%
\begin{equation}
\eta^{ij}D_{jg}^{e}\pi_{ei}=\eta^{ij}\partial_{j}\pi_{gi}+\eta^{ij}%
f_{g~e}^{~a}x_{j}^{e}\pi_{ai}=0. \label{cond1aQ2}%
\end{equation}

$dE$ can be constructed,%
\begin{align}
dE  &  =\left\{  E,\phi^{t}\right\}  dt+\left\{  E,\phi_{g\left(  v\right)
}^{0}\right\}  dv_{0}^{g}+\left\{  E,\phi_{g\left(  v\right)  }^{i}\right\}
dv_{i}^{g}+\nonumber\\
&  -\int dx\int dw\left(
\begin{array}
[c]{cc}%
\left\{  E,\phi_{g\left(  x\right)  }^{i}\left(  x\right)  \right\}  &
\left\{  E,\phi_{i}^{g\left(  \pi\right)  }\left(  x\right)  \right\}
\end{array}
\right)  .\nonumber\\
&  .\left(
\begin{array}
[c]{c}%
-v_{i}^{g}\left(  w\right)  \delta\left(  w-x\right) \\
-\frac{\delta L^{v}\left(  w\right)  }{\delta x_{i}^{g}\left(  x\right)  }%
\end{array}
\right)  dt\left(  w\right)  , \label{dEQ2}%
\end{align}
and the equations of motion are stablished:%
\begin{equation}
dx_{g}^{k}=\left(  -\eta_{00}\pi_{g}^{k}+\partial^{k}v_{g0}-f_{dge}v_{0}%
^{d}x^{ek}\right)  dt, \label{dxiQ2}%
\end{equation}
\
\begin{align}
d\pi_{g}^{i}\left(  y\right)   &  =\left[  \partial_{k}F_{g\left(  x\right)
}^{ki}-F_{a\left(  x\right)  }^{ik}f_{g~b}^{~a}x_{k}^{b}+\right. \nonumber\\
&  -\eta^{00}\eta^{ki}f_{dag}v_{k}^{a}v_{0}^{d}+\eta^{00}\eta^{ki}f_{bag}%
v_{0}^{b}\partial_{k}v_{0}^{a}+\nonumber\\
&  \left.  -\eta^{00}\eta^{ij}f_{b\;g}^{\;a}f_{dac}v_{0}^{b}v_{0}^{d}x_{j}%
^{c}\right]  dt\left(  w\right)  . \label{dpiiQ2}%
\end{align}
With (\ref{cond2Q2}) it follows%
\[
d\pi_{g}^{i}=\left(  D_{kg}^{e}F_{e\left(  x\right)  }^{ki}+f_{b~g}^{~a}%
v_{0}^{b}\pi_{a}^{i}\right)  dt.
\]

One can note that, for QCD in both approaches, $v_{g}^{k}$\ have been
determined by integrability conditions. In the degenerate approach, the
non-abelian generalization of the Gauss law, $D_{ja}^{e}\pi_{e}^{j}=0$, arises
as an integrability condition, while in the standard one it can be obtained
only when equations of motion and a consistency condition ($\pi_{a}%
^{0}=0\Rightarrow d\pi_{a}^{0}=0$) are applied.

\section{Final Remarks}

In this work we could see how the first order Hamilton-Jacobi approach can be
used to describe systems with higher order derivatives. With an extension of
the configuration space, we were able to make two different approaches for
such systems in analogy to what is known in the hamiltonian formalism (SA and
DC). As it is seen in this work, in the HJ context we see that all results
obtained in the SA are obtained in the DC description. However it is important
to notice that it was possible only when a kind of consistency condition was
used in the SA. This seems to be an advantage of the DC approach, where no
extra condition must be considered and therefore less calculations must be made.

One interesting feature of the application of the first order HJ formalism to
DC approach is that no redefinition of the singularity of the theory had to be
made, as it happens in \cite{Degen}. According to Gitman and Tyutin, the gauge
character of the theory would strictly correlated to the singularity of the
new Hessian matrix proposed by them when degenerate coordinates exist. This is
an analysis that cannot be made in this work, since no redefinition of the
Hessian matrix has been done here. However, what one can see in the example of
the QCD (which is a gauge theory with degenerate coordinates) is that, in the
HJ first order approach, perhaps this redefinition is not necessary.

\bigskip

\textbf{Acknowledgements}

BMP would like to thank D. M. Gitman for providing the reference
\cite{GitTyuLya} and BMP and PJP would like to thank D. M. Gitman for
introducing them to the reference \cite{Degen} and the discussions about it.

BMP was partially supported by CNPq and FAPESP; MCB was supported by CAPES;
PJP would like to thank CTA staff for incentive.

\end{document}